# Error Estimation in Large Spreadsheets using Bayesian Statistics

Leslie Bradley, Kevin McDaid
Dundalk Institute of Technology,
Dundalk, Ireland
leslie.bradley@dkit.ie, kevin.mcdaid@dkit.ie

**ABSTRACT**

*Spreadsheets are ubiquitous in business with the financial sector particularly heavily reliant on the technology. It is known that the level of spreadsheet error can be high and that it is often necessary to review spreadsheets based on a structured methodology which includes a cell by cell examination of the spreadsheet. This paper outlines the early research that has been carried out into the use of Bayesian Statistical methods to estimate the level of error in large spreadsheets during cell be cell examination based on expert knowledge and partial spreadsheet test data. The estimate can aid in the decision as to the quality of the spreadsheet and the necessity to conduct further testing or not.*

## 1. INTRODUCTION

Spreadsheets today are used in many industries throughout the world for many different purposes such as decision making, budget forecasting, corporate expansion, and risk analysis to name a few. Spreadsheet technology is a fundamental reporting and decision making tool in most sectors of business. Some areas like the financial sector employ small armies of people who use spreadsheets on a daily basis. In the city of London, [Croll, 2005], one interviewee stated in reference to the finance sector that,

*"Spreadsheets are integral to the function and operation of the global financial system".*

The Sarbanes-Oxley Act of 2002 has forced companies to examine the role spreadsheets play in their financial reporting and decision-making processes. This can require an increase in company resources to adhere to such regulations. There have been many real world examples of spreadsheet errors, loan amounts in a spreadsheet from the Agricultural Bank of Namibia varied from N$59.5 million in one spreadsheet to N$50.4 million in another, [EuSpRiG, 2008].

Spreadsheets continue to be used even though spreadsheet developers use little or no standard development practices from the software industry. This has given rise to a high level of error in spreadsheets. Research into the impact of errors, [Powell, 2007b], discovered that the impact of errors varies depending on the organisation. For 25 operational spreadsheets, ten spreadsheets showed an error impact ranging from $216,806 to $110,543,305; six spreadsheets had errors with no impact while nine had no errors.

An important element of spreadsheet development is testing, where spreadsheets are examined by professionals to ensure their correctness. In the city of London, [Croll, 2005], reports that financial modelling was being limited by the 256 column limit in Microsoft Excel. The author also reports that spreadsheets greater than 1GB in size are already in existence. Thus, to examine each cell can be very time consuming due to their size.

This paper investigates how the overall error rate of a spreadsheet can be inferred from a combination of expert knowledge or belief and partial test result information. This information can be used to decide if further cell by cell testing is required or even whether the spreadsheet has such a high level of error that it should be redeveloped. The methods are ideally suited to large spreadsheets.

The layout of the paper is as follows. Section 2 details aspects of the research into spreadsheet error. In Section 3 the paper describes the problem the research is attempting to address. The research questions to address this problem are presented in Section 4. The statistical model for error rate estimation is described in Section 5 and Section 6 explains how the model is applied to real spreadsheets with a focus on suitable measures for decision making.

## 2. SPREADSHEET ERROR RESEARCH

Reduction of errors in spreadsheets has been the focus for many people in the research community. There have been various investigations into possible influences of error in spreadsheets.

### 2.1 Human Impact

One approach taken is to study the human impact on spreadsheets. Baker, et al [Baker, 2006], conducted a questionnaire online to investigate the developer's competence in creating/editing spreadsheets. They discovered that spreadsheet practices differ greatly from person to person and organisation to organisation. In [Benham, 2005], the authors attempt to reduce the number of errors in spreadsheets by introducing a training technique. They believe that if users have more "error awareness" that they are less likely to make errors.

Bishop and McDaid [Bishop, 2008] looked at the possibility that the level of experience in spreadsheet design can influence the user's performance and behaviour in debugging spreadsheets. In [Panko, 1996, 1998], the author suggests that when humans do simple tasks like typing they can make undetected errors in 0.5% of all actions. This rises to 5% when doing more complex activities like writing a program or a paragraph of text. Thorne and Ball, [Thorne, 2005] look at human factors and their impact in spreadsheet development.

### 2.2 Spreadsheet Reviewing

Another approach used by researchers is to develop a spreadsheet reviewing methodology to detect errors. This method can often involve checking the spreadsheet manually, with the use of auditing tools or a combination of both. Powell, [Powell, 2008] developed an audit protocol that combined different techniques such as mapping and formula checking, with two auditing tools, Xlanalyst and Spreadsheet Professional.

[Panko, 2006], suggests that numerous testing practices could be used on spreadsheets by adopting software checking techniques. Panko suggests that the most extensive way of detecting errors is to conduct a cell by cell search of the spreadsheet, [Panko, 2000]. Croll, [Croll, 2003], reports on the common reviewing process for spreadsheets in the City of London. This entails different levels of reviewing combined with the documentation of each stage.

The HM Revenue & Customs use the "SpACE" audit methodology to check spreadsheets, [HM Revenue, 2007]. The methodology contains 5 steps to audit the spreadsheet including an early decision on whether the impact of the spreadsheet justifies a complete inspection. In step three, a decision on whether the testing should begin is made by the auditor depending on factors outside of the spreadsheet such as the developer, organisation policy, problem complexity, etc.

### 2.3 Error Rates

Various researchers have investigated the error rates for spreadsheets. Panko, [Panko, 2005], looked at 13 studies involving operational spreadsheets that were conducted over a period of nine years, 1995 – 2004. He found an average cell error rate (CER) of 5.2% for 43 of the spreadsheets reviewed in the studies. The CER is the percentage of cells that contain errors. The results show a spreadsheet error rate (SER) of 94% for the 88 spreadsheets studied, i.e. 94% of the 88 spreadsheets contained at least one error.

A later study, [Powell, 2007a], examined 50 operational spreadsheets from various sources, such as a bank, a college and two consulting companies to name a few. The authors measured the errors in terms of error cells and instances. An instance is the occurrence of error in the spreadsheet. An instance usually involves more than one cell with the results showing an instance on average involving 10.05 cells. They discovered a much lower CER of 1.79% over 270,722 formulas.

The authors split the results into two categories; wrong result and poor practice. The wrong result section represents the quantitative errors while poor practice represents qualitative errors. For example, suppose a formula contains a hard coded number. If the result is incorrect, the cell would be deemed a wrong result. If the result is correct then the cell is judged to be a poor practice error. The CER reduced to 0.87% for wrong results, which is significant but lower than previously thought and much lower than 5.2%. Nevertheless, there is a well-accepted issue with the level of error in spreadsheets and Panko [Panko, 2007] summarises it well when he states that "the issue is not whether there is an error but how many errors there are and how serious they are".

### 2.4 Issues with the Error Rates

The cell error results in [Powell, 2007a] are based on formula cells only. The results in [Panko, 2005] are based on 13 studies with the individual CER's based on different measures. One such CER is based on high risk formulas only which do not give an accurate CER for the overall spreadsheet.

### 2.5 Rationale

The goal of spreadsheet reviewing is to establish its correctness. This can be almost impossible for large spreadsheets with millions of cells. Croll has reported on the size of spreadsheets in the City of London but the number of unique formulas can range from 1,000 to 10,000 and upwards, [Croll, 2003]. Instead we propose a method whereby expert information or belief as to the likely cell error rate can be combined with partial test information to infer an error rate for the remaining untested cells.

Powell, [Powell, 2007b], found 9 out of 25 spreadsheets contained no errors. Five of these error free spreadsheets came from the same organisation. Therefore it is reasonable to infer a belief that a further spreadsheet provided by this organisation will have a lower than average cell error rate. Similarly, knowledge of the developer's ability and the spreadsheet's complexity will influence to some extent the reviewer's belief as to the likely cell error rate. This is important information that can be used to hone any estimate of the cell error rate. Similarly, judgements of likely quality based on the presence or absence of good practices and controls such as documentation, version control and design patterns amongst others can influence an expert's judgement as to the likely error rate.

We feel strongly that this belief should be incorporated into the prediction model, and combined with available test information. The natural approach is a Bayesian one which we introduce in the following subsection.

### 2.5 Software Reliability

Research into software reliability has looked at the application of Bayesian networks to predict software reliability with limited or uncertain data. The use of Bayesian statistics allows the combination of different types of information available to the developer, [Fenton, 2004, 2007b]. The use of Bayesian networks can include qualitative information e.g. staff experience and testing quality, [Cockram, 2001; Fenton, 2007b; 2007a].

Researchers using Bayesian methods to predict software reliability, [Fenton, 2007b] suggest "that a model constructed using expert judgement and historical data, within one organisation can be used within the same organisation to predict accurately the outcome of new projects." This research is looking at the use of Bayesian methods to combine expert information with test data to give improved prediction of cell and spreadsheet error rates during review.

## 3. PROBLEM

Spreadsheet audits can require huge amounts of time and resources to conduct as mentioned by [Croll, 2003], "The time taken to review these models can range from twenty five hours to many hundreds, generating significant fee income for firms undertaking this work,". If the level of error in the spreadsheet could be predicted, then a decision on whether the spreadsheets should be tested could be made. This would focus resources on the more error prone spreadsheets.

Information on the likely cell and spreadsheet error rate could be of great importance to management, allowing them to make a decision on whether a spreadsheet requires exhaustive testing or, indeed, whether it should be rejected and redeveloped. The prediction of likely cell and spreadsheet error rate, by an auditor, say, can be helped by expert and other knowledge such as developer experience or available audit information. The auditor can use this to aid the decision about the spreadsheet.

## 4. RESEARCH QUESTIONS

This work is part of a larger study and the particular question that we are trying to solve is can a model be established to predict the CER of large spreadsheets based on both expert knowledge and available test data. The predicted CER can then guide the decisions based on cell error rates.

The research first looked at literature on existing spreadsheet error research and general statistical methods for reliability including Bayesian methods. The relationship between the CER and the SER was investigated using statistical methods. This work was examined though the following questions:

> **RQ1**: What does existing research say about the level of spreadsheet error and methods to discover errors in spreadsheets?

> **RQ2**: What statistical methods can be used to predict cell and spreadsheet error rates?

The research completed for **RQ1** has looked at existing research into the level of errors in spreadsheets. This is described in Section 2. The research completed for **RQ2** first examined the relationship between the CER and the SER. A key issue in this regard is the nature of the dependence, if any, between cell errors. The research looks at defects, which are errors in formulas that produce an incorrect result. A dependence structure indicates that the probability of one cell containing an error influences the probability of any other cell containing an error whereas an independence structure assumes that no relationship between the error cells exists. In this work we make the critical assumption that unique cell formula errors are independent and we discuss this later the conclusion.

The research has also shown there are many different sources of information on spreadsheet error. The investigation aims to combine both prior knowledge and test data to estimate the CER of large spreadsheets. The possibility of creating a structure is explored through **RQ3**.

> **RQ3**: Can a model be developed that combines prior knowledge and available test data to estimate the CER for large spreadsheets?

We propose the use of Bayesian methods that combine prior information and test data to answer **RQ3**. There are many external factors that can influence the CER of a spreadsheet. The developer and the organisation can impact the CER at the early stage of spreadsheet development. The complexity of the spreadsheet can naturally influence the error rate. The more complex the spreadsheet then the more likely it is to contain an error. The model will be evaluated using large spreadsheets once completed. The model will be tested to predict the CER with the results compared to the actual CER of the spreadsheets.

> **RQ4**: How effective is the model at predicting the cell error rate in spreadsheets?

The fourth research question will be addressed in future work. The model will be evaluated using operational spreadsheets. The validity of the assumptions underlying the model will also be evaluated.

## 5. THE BAYESIAN MODEL

The model aims to estimate the CER of large spreadsheets by combining expert information and information on the correctness of individual cells gathered during partial review. The expert information is known as “prior information”. The spreadsheet may be partially examined. This information is called “test data”. Both sets of information can be combined using Bayesian methods to produce “posterior information”, from which error rate estimates can be derived.

### 5.1 Prior Information

The prior information can be based on a number of factors that influence the CER of a spreadsheet; the company, the developer, the spreadsheet complexity, etc. The characteristics of the developer and the company regulations play an early role in the life of a spreadsheet. The company environment could contribute to the CER as deadlines need to be met which can replace quality with speed. The developer may also not have the relevant skills or experience to complete the task which can produce more errors than an experienced developer. The spreadsheet complexity represents the level of complex calculations, formulas, and functions in similar/previous spreadsheets. These factors could indicate the level of error in the spreadsheet.

Further qualitative information based on initial structural and content investigation, often an initial step in a review methodology, can also influence the choice of prior distribution.

The prior information is represented as a beta distribution, the form of which is given below. This is the conjugate prior of the binomial distribution which represents the test data in section 5.2. The beta distribution for the cell error rate, $\theta$, has two parameters, _ and _. These are set by the individual or individuals with best knowledge of this or related spreadsheets and represent the belief as to its likelihood for each of the possible cell error rate values. This expert may well be the reviewer who has already performed a cursory examination of the spreadsheet. In cases where very little is known about the spreadsheet a vague prior can be chosen which will reflect this lack of certainty regarding the likely error rate.

$$f(\theta) = \frac{\theta^{\alpha-1}(1-\theta)^{\beta-1}}{\int_{\theta} \theta^{\alpha-1}(1-\theta)^{\beta-1}} \quad \text{where } 0 \le \theta \le 1$$

*The Beta Distribution*

The mean and variance of the distribution are as follows:

$$Mean = \frac{\alpha}{\alpha+\beta} \quad \text{and} \quad Variance = \frac{\alpha\beta}{(\alpha+\beta)^2(\alpha+\beta+1)}$$

The prior beta distribution can be used to produce an estimate of the likely CER before review begins based only on expert information which can be used to aid the decision as to whether detailed cell by cell testing is required. If the prior CER is not acceptable, then a decision to review the spreadsheet can be made with the possibility of further decision steps.

**5.2 Test Data**

The test data is information obtained by reviewing the cells in the spreadsheet individually. A binomial distribution can be used to model the occurrence of failures, errors, during this review. The binomial distribution returns the probability distribution of the number of x successes (number of errors) in n independent trials (number of tested cells) with probability $\theta$ of a success. The independence assumption addressed in section 6.1 is important,

$$b(x;n,\theta) = \binom{n}{x}\theta^{x}(1-\theta)^{n-x} \text{ for } x = 0,1,2,\ldots,n$$

*The Binomial Distribution*

### 5.3 Posterior Information

The posterior information combines the prior information (expressed through a beta distribution) and the test data (modelled by a binomial distribution) to produce a statistical distribution which represents the likely values for the error rate for the cells that have not yet been tested. This is known as the posterior distribution. Using a beta prior distribution and a binomial model the posterior distribution also takes the form of a beta distribution, beta(_+x, _+(n-x)), the distribution for which is presented below.

$$f(\theta) = \frac{\theta^{\alpha+x-1}(1-\theta)^{\beta+(n-x)-1}}{\int_{\theta}\theta^{\alpha+x-1}(1-\theta)^{\beta+(n-x)-1}} \quad \text{where } 0 \le \theta \le 1$$

*The Posterior information in the form of the Beta Distribution*

The posterior CER distribution gives an updated estimate of the level of error in the spreadsheet and can be updated as more test data is obtained.

### 5.4 Beta Distribution of Prior and Posterior Information

The following section illustrates prior and posterior distributions. Suppose a spreadsheet contains 900 formula cells and expert information indicates a cell error rate of 0.2 with a Std. Dev. of 0.0146. This yields parameter values of 2 and 8 for the prior beta distribution. This information could also be considered as equating to having knowledge of 10 cells in the spreadsheet and finding 2 of these with errors.

Assume further that the first 20 formula cells are tested and the results found 2 error cells. This test data is combined with the prior, beta (2, 8), to produce a posterior beta (4, 26). The mean of the prior distribution is 0.2 with the mean of posterior distribution 0.13.

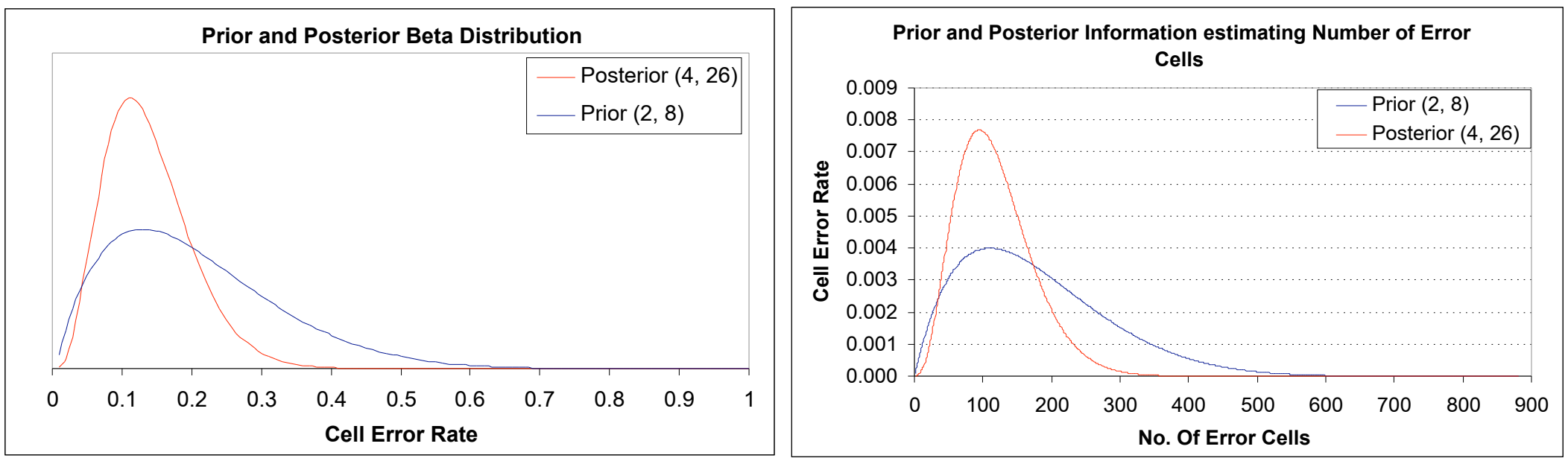

**Figure 1(a) Prior and Posterior information in terms of Cell Error Rate and (b) Prior and Posterior probability estimates for number of error cells.**

In Figure 1(a), the beta distribution is shown using prior information. The distribution suggests the likely CER is approx. 0.13 or 13%. The posterior information suggests the likely CER is approx 0.12 or 12%. This is almost double the prior's probability. The shape of the distribution has now changed from being wide around the mean for the prior to taller and thinner for the posterior. This shows our belief that the posterior information is more certain then the prior information. A narrower prior distribution represents a stronger belief in the cell error rate.

Figure 1(b) shows the likely number of error cells of the remaining 880 untested ones based on both prior and posterior distributions. The prior distribution shows that the number of errors is likely to range from 0 to 550 approximately with 110 the most likely number of error cells. The posterior distribution indicates a smaller range of 0 to 320 likely errors approximately 96 errors cells.

## 6. POSTERIOR INFORMATION

This section illustrates the methodology through its application to three operational financial spreadsheets. The spreadsheets relate to research project finance in a third level institute. The spreadsheets were audited by the authors using the [Powell, 2008] methodology. The analysis was conducted by two people independently. The results were compared and agreement reached over the errors found.

The sequential analysis of the formulas begins in the first worksheet in the first unique formula block. This block is checked for errors and the result is recorded. The process continues onto the next formula, column by column and then row by row in a sequential process. An example of how the data used from the error analysis is shown in Table 1.

| Cells Tested | 1 | 2 | 3 | 4 | 5 | 6 | 7 |
|---|---|---|---|---|---|---|---|
| No. Errors | 0 | 0 | 0 | 1 | 1 | 2 | 3 |
| Mean | 5/101 | 5/102 | 5/103 | 5/104 | 5/105 | 6/106 | 6/107 |

**Table 1: Example of Results**

As formula cells are tested, the numbers of errors are recorded. Assuming a prior with parameter values of (5, 95) is selected beforehand, the mean of the posterior error rate distribution is updated as each unique formula block is checked. The mean is calculated using the formula in section 5.1 and it decreases as no errors are found and increases for discovered errors. Each time a unique formula block is checked a new posterior mean is calculated.

### 6.1 Unique Formulas

Given that the standard measure of error rate is based on cells, we could base our model on individual cells rather than blocks of copied cells. However, the model makes the important assumption that the relationship between error cells is independent. However copied formulas contradict this assumption as the original cell would influence the error status of the copied cells. Therefore the analysis looks at unique formulas in spreadsheets.

| | A | B | C | D |
|---|---|---|---|---|
| 1 | 1 | 5 | | 7 |
| 2 | 3 | 7 | | 8 |
| 3 | 5 | 9 | | 9 |
| 4 | | | | |
| 5 | =A3+A2-A1 | =B3+B2-B1 | | =D3+D2-D1 |

**Figure 2: Formula view of a spreadsheet section**

A spreadsheet section which contains 12 non-empty cells; 3 formula cells and 9 data input cells can be seen in Figure 2. Two of the formulas are copies, cell B5 and cell D5, of the formula in A5. The first unique formula is A5 and can involve more than one cell, if the formula cells are directly above, below, in front or behind with no blank or different value cell between. In Figure 2, there are 2 unique formulas cells, A5 and D5. This is different to root formulas which are measured throughout the spreadsheet, [Panko, 2007].

## 6.2 Applying to Operational Spreadsheets; Defects.

This section presents the analysis of three operational spreadsheets focusing on a distribution for the level of error remaining in the spreadsheet. The results are in terms of unique formulas and with focus on defect errors. The following figures show the best estimate for the defect error rate (through the mean of the posterior distribution) as testing proceeds including 5% and 95% limits which indicate how wide the distribution for the error rate extends. All the graphs are produced based on prior parameters of (5, 95).

Reliability graphs are also included which gives the probability that the error rate for the remainder of the spreadsheet is below a selected acceptable error rate chosen in this case to be 5%. A firm may decide that they will stop the review if they are happy that this acceptable defect rate has not been breached.

### Spreadsheet I

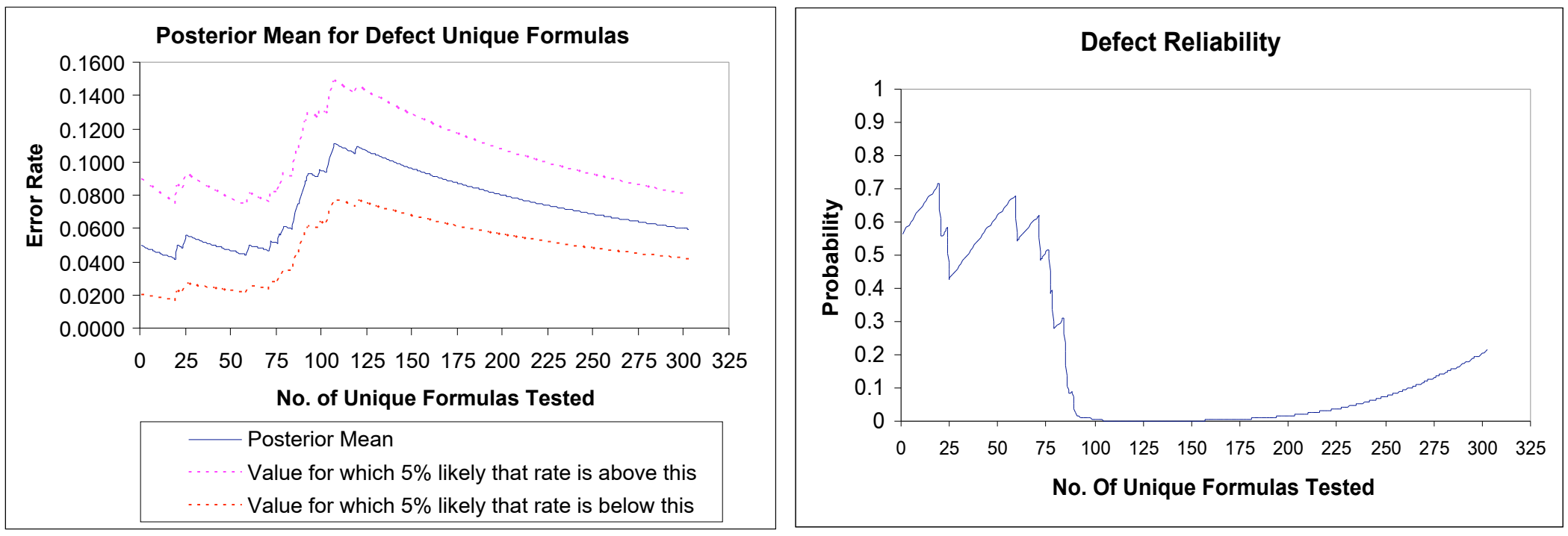

**Figure 3(A) The posterior mean for defects with confindence interval. (B) Defect reliability.**

In Figure 3(A), the posterior mean shows periods where unique formulas contain defects. These periods are shown by the sharp jumps in the mean. The defect free unique formulas are shown by the gradual descents in the graph. The graph also contains two lines either side of the mean. The top curve represents the 5% chance that the CER is greater than this line. The bottom curve gives the error rate for which there is a 5% chance of the true value being below. There is a 90% chance that the true error rate lies between the 5% and 95% curves. The areas between these outside lines; which include the mean, imply 90% confidence that the CER lies in this interval.

Some organisations may allow a certain level of error to remain in the spreadsheet. If the actual error rate falls below this acceptable CER, then the organisation may not deem the spreadsheet to need further examination. The reliability calculates the likelihood of the posterior CER being equal to or below the acceptable CER. In Figure 3(B), the reliability that the defect error rate stays less than or equal to the acceptable CER of 5% sharply decreases with some areas of improvement. The reliability spikes sharply and decreases very closely to 0% for 50 – 100 unique formulas. The reliability begins to increase towards the end and finishes with 20% reliability. This suggests the spreadsheet requires further testing or redevelopment as the reliability is very poor and management cannot be sufficiently sure that the error rate is below an acceptable level. The same reasoning can be applied to the following Figures.

### Spreadsheet II

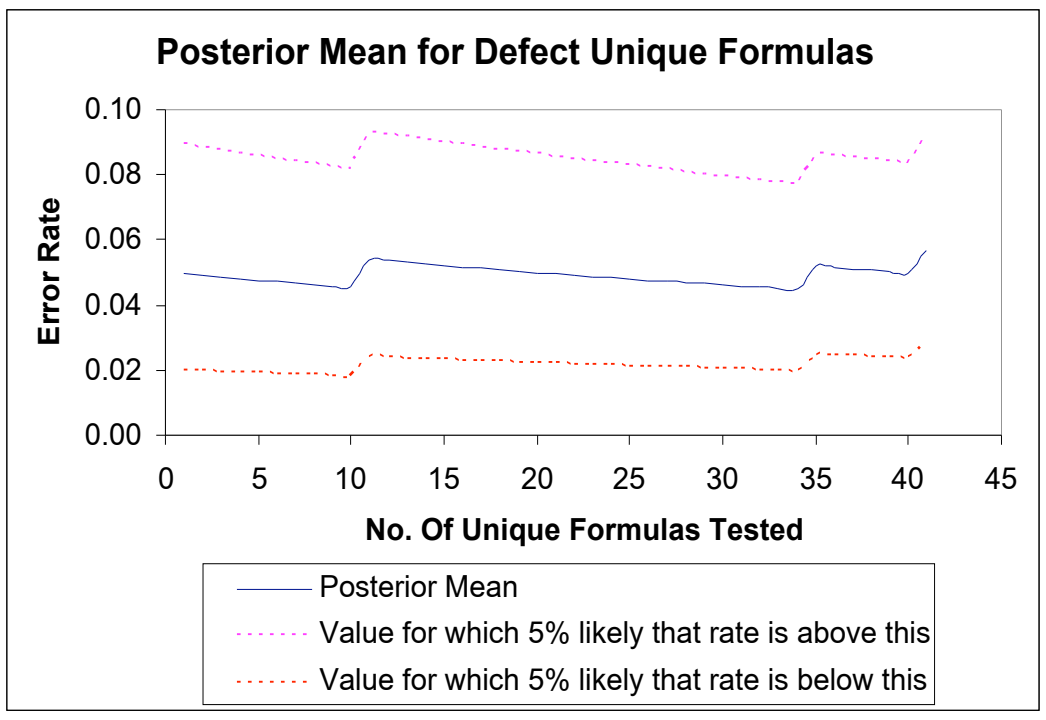

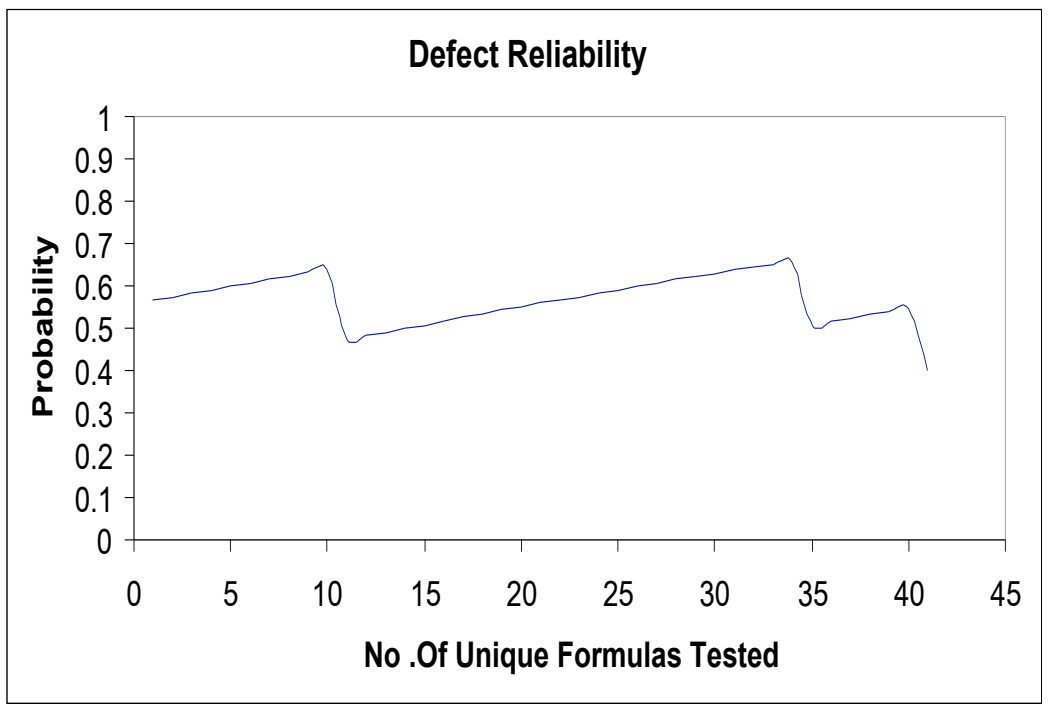

**Figure 4: (A) The posterior mean for defects with confidence interval. (B) Defect reliability**

### Spreadsheet III

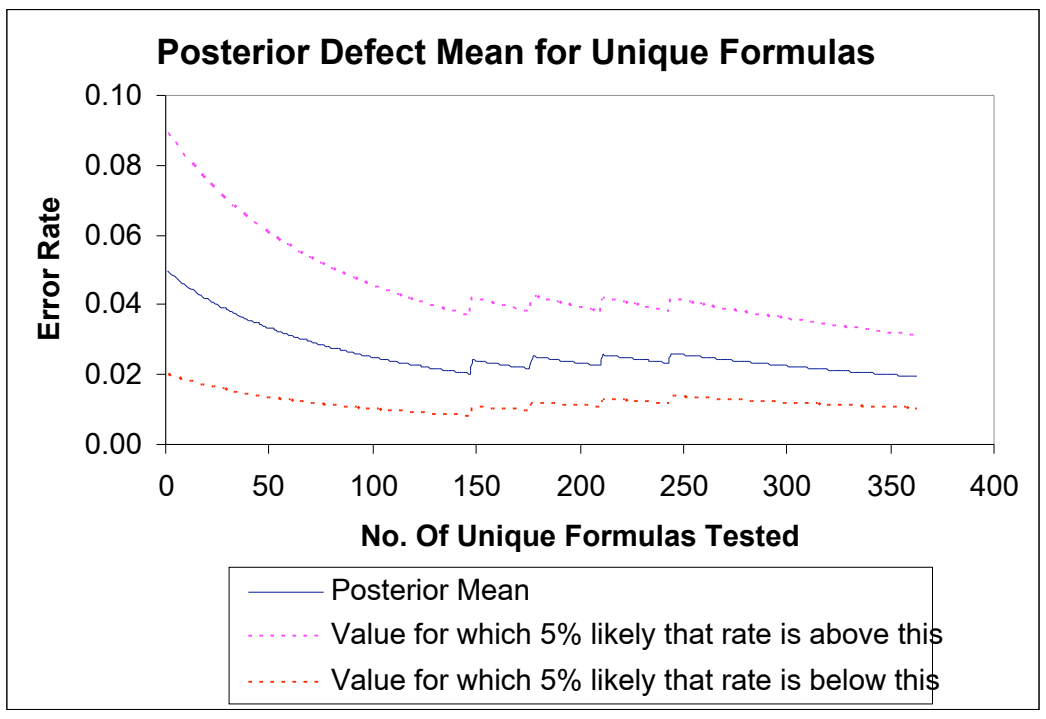

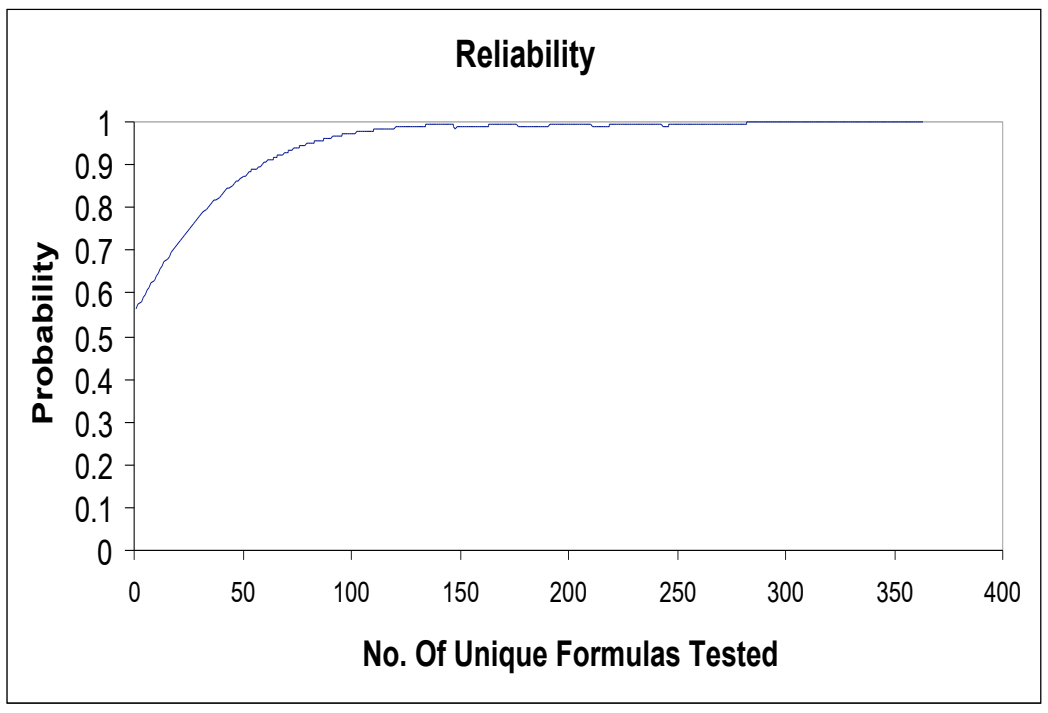

**Figure 5: (A) The posterior mean for defects with confidence interval. (B) Defect reliability**

In Figure 5(A), the posterior defect mean starts at the prior information and reduces until defects are found. The posterior information improves the prior information as the mean reduces and becomes stable to produce conclusions. If the prior information was not included we would see a zero mean until defects were found and would create large jumps. This would prove difficult to produce any conclusions based on the information as it is not stable. In Figure 5(B), the reliability increases and stays above 95% from 100 unique formulas until the end of the spreadsheet. This would indicate after, say 100 – 130 unique formulas have been tested, that testing can stop based on the defined criterion. We believe this to be true by applying this method to improve spreadsheets.

## 7. CONCLUSION

This paper outlines research that examines the use of Bayesian methods to combine test result data with expert knowledge on the likely error rate to yield posterior estimates of reliability that can support managers in their decision as to how much of a spreadsheet needs to be tested. This method may be particularly suited to large spreadsheets for which it can be very costly and time consuming to review the entire spreadsheet.

There are a number of issues with the method. Firstly, the method assumes perfect discovery with respect to spreadsheet error during the review, although research has shown this to be extremely difficult. Secondly, the method does not factor in omission errors into the prediction. Thirdly, we assume that errors are independent. While we record errors based on unique formulas to avoid the clear issue of dependence between copied cells, there are other ways in which cell errors may be independent. Cells that contain the same or related functions may be related in terms of likelihood of error. Likewise, cells in close proximity in a spreadsheet may not be independent in terms of likelihood of error. This can raise questions as to how well review results for one part of a spreadsheet can be used to predict the reliability of another part of the same or a different spreadsheet. A more natural way of achieving better representative coverage of the spreadsheet and to over come this issue might be to randomly choose the formulas to check. This is the subject of ongoing research.

Finally, error rates are only one aspect of spreadsheet quality measurement. As Powell states "obtaining reliable estimates of cell error rate is only a first step toward understanding the fundamental question of the impact that spreadsheets have on the quality of decision making in organisations", [Powell, 2007a].

## 8. Acknowledgements

This work is supported by the Technological Sector Research measure funded under the NDP and ERDF.